\documentclass[12pt,preprint]{emulateapj}

\usepackage{graphicx}
\usepackage{amsmath}
\usepackage{hyperref}
\usepackage{color}
\usepackage{bm}

\hypersetup{
    colorlinks=true,
    linkcolor=red,
    citecolor=blue,
    urlcolor=magenta,
}

\begin{document}

\title{Chromatic microlensing time delays}
\author{Kai Liao$^{1}$}\email{liaokai@whut.edu.cn}

\affil{
$^1$ {School of Science, Wuhan University of Technology, Wuhan 430070, China}
}

\begin{abstract}
  Due to the finite size of the disk and the temperature fluctuations producing the variability,
  microlensing changes the actual time delays between images of strongly lensed AGN on the $\sim$day(s) light-crossing time scale of the emission region.
  This microlensing-induced time delay depends on the disk model, primarily the disk size $R_\mathrm{disk}$ which has been found to be larger than predicted by the thin-disk model.
  In this work, we propose that light curves measured in different bands will give different time delays since $R_\mathrm{disk}$ is a function of wavelength,
  and by measuring the time delay differences between bands, one can 1) directly verify such an new effect; 2) test the thin-disk model of quasars.
  For the second goal, our method can avoid the potential inconsistency between multi-band light curves that may bias the results by continuum reverberation mapping.
  We conduct a simulation based
  on a PG 1115+080-like lensed quasar, calculating the theoretical distributions of
  time delay differences between two bands: u and i centered around 354nm and 780nm, under and beyond the thin-disk model, respectively.
  Assuming the disk size is twice larger than the standard one, we find that with a precision of 2 days in the time delay difference measurements,
  the microlensing time delay effect can be verified with $\sim4$ measurements while with $\sim35$ measurements the standard model can be excluded.
  This approach could be realized in the ongoing and upcoming multi-band wide-field surveys with follow-up observations.

\end{abstract}


\section{Introduction}
Quasars are excellent objects for studying black hole physics, galaxy formation and cosmology.
The accretion disk of the central supermassive black hole has typical size of $\sim10^{15}$cm
and generates fluctuations with time scale $\sim$ days of the light curve.
Theoretically, a standard, non-relativistic, thin-disk model emitting as a blackbody~\citep{Shakura1973} is usually taken to describe the disk.
In addition, the variability is described by the ``lamp post" model~\citep{Cackett2007} in which the fractional temperature variation
is independent of radius in the disk with a time lag determined by the light travel time from the disk center.

Currently, neither ground- or space-based telescopes can resolve quasar accretion disks. To estimate the size of emitting region,
techniques such as reverberation mapping of the quasar continuum emission~\citep{Collier1998,Sergeev2005,Shappee2014,Fausnaugh2016}, modeling of the structure of the broad Fe K$\alpha$ line~\citep{Tanaka1995,Iwasawa1996,Fabian2002,Iwasawa2004}
and quasar microlensing~\citep{Pooley2007,Morgan2008,Blackburne2011,Chartas2012,Mosquera2013,MacLeod2015} have been well-developed.
For the reverberation mapping method, an adequate summary of current works is that light curves in different bands are highly correlated and well-modeled by the damped random
walk (DRW) such that methods like interpolated cross-correlation function~\citep{Peterson1998} or JAVELIN~\citep{Zu2013} are able to capture the appropriate lags.
When the brightness in the red increases, an even greater increase in the brightness in the blue trends to be seen - but the connection does not
seem to be fully deterministic. In other word, light curves in different bands are only partially
coherent (see~\citep{Sun2014} and references therein) and a skewed transfer function could bias the estimates~\citep{Chan2020}.

For the microlensing method,
the microlensing light curves are generated on top of the magnifications by strong lensing.
A large number of strongly lensed quasars have been and will be observed and well-studied~\citep{Treu2010,Oguri2010}.
Stars in the lens galaxies as the secondary lenses bring extra magnifications known as the microlensing effect.
Depending on the local environments of the images and peculiar motion of the source relative to the lens and the observer,
the magnifications change in the form of microlensing light curves. A larger disk size relative to the mean Einstein
radius of the stars will smooth out the microlensing effect and result in smaller variability amplitudes.
The measured amplitude of microlensing variability constraints the size of the disk in turn.
In recent years, evidences from microlensing together with reverberation mapping have implied that not all quasars are well
modeled by the thin-disk model, nor by the lamp-post model of variability.
The measured disk sizes seem to be larger than those predicted by the thin-disk model~\citep{Morgan2010,Rojas2014,Edelson2015,Lira2015,Fausnaugh2016,Jiang2017}.

Another interesting phenomenon caused by microlensing is the time delay changing between lensed AGN images, recently proposed by~\citep{TK18}.
The measured time delays based on the light curve pairs which suffer from the contamination of microlensing magnification
patterns~\footnote{Such measurements were proved to be accurate. See the COSMOGRAIL program: www.cosmograil.org and the Time Delay Challenge (TDC) program~\citep{Liao2015}.} were proposed not
to be the cosmological one~\citep{TK18}. Microlensing produces changes in the actual time delays on the $\sim$day(s) light-crossing time scale of the emission region
due to a combination of the inclination of the disc relative to the line of sight and the differential magnification of the temperature
fluctuations producing the variability.
The microlensing-induced time delay will slowly change as the accretion disk moves relative to the stars doing the microlensing~\citep{TK18}.
This effect (if it exists) would bias the time delay measurements and then the Hubble constant ($H_0$) inference, especially for small time delays~\citep{TK18,Liao2019}.
In principle, one can potentially verify such an effect using the variation of the time delays measured at different epoches/seasons of the light curves in single waveband. However,
due to the uncertain peculia motion and the source position, it may require time scales much larger than that
typically gives a time delay measurement $\sim$ year(s) to observe
the variation. Current studies have not found the variation~\citep{Bonvin2018,Chen2018} and the effect was not considered in cosmological studies~\citep{Wong2020}.
Therefore, directly verify/measure the effect and its details would benefit understandings of quasar physics, microlensing and cosmology.

In this work, we propose to measure time delay differences between bands on a single monitoring campaign. Such measurements
can not only verify this new microlensing phenomenon but also test the thin-disk model by
measuring the disk size within a short timescale.
The microlensing time delay effect primarily depends on the disk size as a function of wavelength~\citep{TK18}.
Therefore, time delays measured in different bands correspond to different disk sizes and are perturbed by microlensing differently.
As a supplement in studying disk size, our method does not directly analyze light curves in different bands as a whole and
can avoid the multi-band light curve inconsistency problem and the details of the transfer function in continuum reverberation mapping ~\citep{Sun2014,Chan2020}.
In absence of microlensing, delays in different bands should be exactly the same whereas the microlensing time delay effect will
introduce different scatters in the time delays measured in different bands.

The Letter is organized as follows. In Section 2, we briefly introduce the thin-disk model of quasars and the microlensing time dely effect;
In Section 3, we propose to measure the chromatic microlensing time delays;
Then we do a simulation to show how our method is viable in Section 4; Finally, we summarize and make discussions in Section 5.

\section{Thin-disk model and microlensing time delays}
In a standard, non-relativistic, thin-disk model that emits as a blackbody, for rest-frame wavelength $\lambda_\mathrm{rest}$,
the characteristic radius of the disk where the temperature matches the photon wavelength $kT=h_\mathrm{p}c/\lambda_\mathrm{rest}$
is given by~\citep{Shakura1973}
\begin{equation}\label{R0}
\begin{aligned}
R_0&=\left[\frac{45G\lambda_\mathrm{rest}^4M_\mathrm{BH}\dot{M}}{16\pi^6h_\mathrm{p}c^2}\right]^{1/3} \\
&=9.7\times10^{15}\left(\frac{\lambda_\mathrm{rest}}{\mathrm{\mu m}}\right)^{4/3}\left(\frac{M_\mathrm{BH}}{10^9M_\odot}\right)^{2/3}\left(\frac{L}{\eta L_\mathrm{E}}\right)^{1/3}\mathrm{cm},
\end{aligned}
\end{equation}
where $k,G,h_p,c$ are Boltzmann, Newtonian, Planck constants and light speed, respectively. $M_\mathrm{BH}$ is the central black hole mass, $L/L_\mathrm{E}$ is luminosity in units of the Eddington luminosity,
$\eta=L/\dot{M}c^2$ is the accretion efficiency which ranges from $\sim0.1$ to $\sim0.4$ positively correlated with the spin of the black hole.
Note that $\eta$ is hard to directly measure. However, $R_0$ with current measurement precision is not sensitive to the adoption of $\eta$ due to
the power exponent -1/3 dependence, i.e., the fact of larger size from observation can not be explained even by choosing the minimum $\eta$.

A dimensionless radius is usually defined as
\begin{equation}\label{xi}
\xi=\frac{hc}{kT_0(R)\lambda}=\left(\frac{R}{R_0}\right)^{3/4}\left(1-\sqrt{\frac{R_\mathrm{in}}{R}}\right)^{-1/4},
\end{equation}
where $R>R_\mathrm{in}=\alpha GM_\mathrm{BH}/c^2$, the inner edge of the disk~\citep{Morgan2010}, and $\alpha=1$ and 6 for a Schwarzschild and Kerr black hole, respectively. The unperturbed temperature profile $T_0(R)^4\propto R^{-3}(1-\sqrt{R_\mathrm{in}/R})$
and the unperturbed surface brightness profile is
\begin{equation}
I_0(R)\propto (e^\xi-1)^{-1}.
\end{equation}

For the variability, the ``lamp post" model~\citep{Cackett2007} is often used, where the fractional temperature variation is independent of $R$:
\begin{equation}
T(R,t)=T_0(R)[1+f(t-R/c)],
\end{equation}
where $f(t-R/c)$ is the lagged fractional luminosity variability since the light travels from the disk center as the driving source.
Therefore, the variable surface brightness can be got by Taylor expanding the blackbody function:
\begin{equation}
\delta I(R,t)\propto f(t-R/c)G(\xi),
\end{equation}
where
\begin{equation}\label{G}
G(\xi)=\frac{\xi e^\xi}{(e^\xi-1)^2}.
\end{equation}

The mean microlensing-induced time lag is given by~\citep{TK18}
\begin{equation}\label{tmicro}
t_\mathrm{micro} = \frac{1+z_s}{c}\frac{\int dudvG(\xi)M(u,v)R(1-\cos\theta \sin \beta)}{\int dudvG(\xi)M(u,v)}-t_\mathrm{disk}
\end{equation}
where $M(u,v)$ is the microlensing magnification map projected in the source plane, $\theta$ is the polar angle in the accretion disk plane,
$\beta$ is the inclination angle with $\beta=0$ corresponding
to a face-on disk. $u=R\cos\theta\cos \beta$ and $v=R\sin\theta$ are the coordinates in the source plane.
Note that we have separated out the mean time lag caused by the disk itself relative to the driving source $f(t)$ corresponding
to the case without microlensing $M(u,v)\equiv1$ (i.e., the so-called \emph{geometric delay} with which one can apply the reverberation mapping):
\begin{equation}\label{diskt}
t_\mathrm{disk}=\frac{1+z_s}{c}\frac{\int dudvG(\xi)R(1-\cos\theta \sin \beta)}{\int dudvG(\xi)}.
\end{equation}
Ignoring the inner edge of the disk, $t_\mathrm{disk}=5.04(1+z_s)R_0/c$, where the time-scale
\begin{equation}
\frac{(1+z_s)R_0}{c}\simeq \frac{3.8\ \mathrm{days}}{(1+z_s)^{1/3}}\left(\frac{\lambda_\mathrm{obs}}{\mathrm{\mu m}}\right)^{4/3}\left(\frac{M_\mathrm{BH}}{10^9M_\odot}\right)^{2/3}\left(\frac{L}{\eta L_\mathrm{E}}\right)^{1/3}.
\end{equation}

\section{Chromatic microlensing time delays}
Both $t_\mathrm{micro}$ and $t_\mathrm{disk}$ primarily depend on the disk size as a function of observed wavelength $\lambda_\mathrm{obs}=(1+z_s)\lambda_\mathrm{rest}$:
$R_0(\lambda_{obs})\propto\lambda_\mathrm{obs}^{4/3}$ following ``the redder, the larger" law.
Note that $t_\mathrm{micro}$ also depends on the observing epoch $\mathcal{T}_\mathrm{obs}$ as well since the source experiences different positions in the
magnification map due to the relative peculia motion.
What one measures are the arrival timings of the light curve phases.
Compared to the phases of driving source $f(t)$, the light curve phases of image $P$ has a total time lag (hereafter we use the terms ``time lag" for single image and ``time delay" between two images, respectively):
\begin{equation}\label{ttotal}
\begin{aligned}
t^P_\mathrm{total}(\lambda_\mathrm{obs},\mathcal{T}_\mathrm{obs})=t_\mathrm{cosm}+t^P_\mathrm{sl}+t^P_\mathrm{sub}+t_\mathrm{disk}(\lambda_\mathrm{obs}) \\
+t^P_\mathrm{micro}(\lambda_\mathrm{obs},\mathcal{T}_\mathrm{obs}),
\end{aligned}
\end{equation}
where $t_\mathrm{cosm}$ is the cosmological light propagating time in the absence of the lens galaxy, $t^P_\mathrm{sl}$ is from strong lensing of macro gravity
field of the lens galaxy, $t^P_\mathrm{sub}$ is the perturbation caused by dark matter substructure.
In addition, there is an extra term caused by line-of-sight density fluctuation (weak lensing) which has been incorporated into $t_\mathrm{cosm}$ or $t^P_\mathrm{sl}$ for simplicity.

In practice, what one can directly measure is the time delay between any two images ($P$ and $Q$) of a lensed AGN by comparing their light curve shifting in time domain in the given band $b$:
\begin{equation}
\Delta t_\mathrm{lc}(b,\mathcal{T}_\mathrm{obs})=\Delta t_\mathrm{sl}+\Delta t_\mathrm{sub}+\Delta t_\mathrm{micro}(b,\mathcal{T}_\mathrm{obs}),
\end{equation}
where the widely concerned one for strong lens time-delay cosmography is $\Delta t_\mathrm{sl}$~\citep{Treu2016} determined by a specific cosmological model with parameters therein, and the lens potential.
Note that in principle unbiased time delays should be measured in single band since the intrinsic light curves are only partially correlated between bands~\citep{Sun2014}.
We can always use the monochromatic central wavelength of a band in calculation since the radial width due to the blackbody emission is significantly
more important than the wavelength spread from a typical broad-band filter~\citep{TK18}.
The last term depends on the disk size, the observed wavelength and the observing epoch as well. The existence of microlensing time delays
could bias the $H_0$ measurements if uncounted especially for those with small time delays~\citep{TK18,Liao2019}. In addition, it degenerates with the time delay perturbation caused by dark matter substructure in the lens galaxy
which is supposed to be a fraction of one day~\citep{Keeton2009}, making probing dark matter substructure impossible with time delay/time delay ratio anomalies unless using transient sources like
repeated fast radio bursts or gravitational waves~\citep{Liao2018}. For the impact of dark matter subhalos on inference of $H_0$, see~\citep{Gilman2020}.

Despite of these, one may conjecture that measuring the time delay anomalies
can be used to verify microlensing time delay effect and test quasar models in turn. At this point, it is like what~\citep{Keeton2009} suggested for probing dark matter substructure. However, no robust result in this respect
has appeared up. The anomalies are hardly to be accurately determined by light curves due to the mass-sheet degeneracy, and uncertainties in the lens radial profile, dark matter substructure and $H_0$,
where the $H_0$ issue is currently annoying the community.
Ratio of time delays seems more promising~\footnote{We give a discussion on using time delay ratio anomalies to test the microlensing time delay effect and quasar models in the appendix.} which is free of the degeneracies except the dark matter substructure that may sub-dominate (a fraction of one day).
Nevertheless, this approach is essentially indirect which may suffer from bias in the lens modelling~\citep{Schneider2013,Birrer2016,Ding2020}.

By contrast, what more robust is to directly measure the changing of $\Delta t_\mathrm{micro}(\lambda_\mathrm{obs},\mathcal{T}_\mathrm{obs})$ as a function of either $\lambda_\mathrm{obs}$ or $\mathcal{T}_\mathrm{obs}$ .
For the latter one, due to the uncertain peculia motion and source position, it may require time scales much larger than that
typically gives a time delay measurement $\sim$ year(s) to observe
the variation (larger transverse peculia motion and positions closer to caustic lines where time delays change rapidly are easier). In addition, it is complicated to
calculate the theoretical distribution of the time delay variation since it requires the prior on
the peculia motion distribution and it entails correlated points or line integrals in the time delay variation maps. Furthermore, the gravity field may
not keep static for very large time scales due to the motion of the dark matter subhalos that could also change the time delays.

Therefore, to verify the existence of microlensing time delays and test quasar models,
we suggest that one measure time delays in two different bands $b1$ and $b2$ at the same epoch. Their difference $\Delta t_\mathrm{lc}(b2,\mathcal{T}_\mathrm{obs})-\Delta t_\mathrm{lc}(b1,\mathcal{T}_\mathrm{obs})$
is determined by the microlensing time delay difference:
\begin{equation}\label{compare}
\delta\Delta t_\mathrm{lc}(b1,b2,\mathcal{T}_\mathrm{obs})=\delta\Delta t_\mathrm{micro}(b1,b2,\mathcal{T}_\mathrm{obs}),
\end{equation}
where the symbols $\delta$ and $\Delta$ are for different bands and different lensed images, respectively. $\delta\Delta t_\mathrm{lc}$ should be independent of the intrinsic light curves and the algorithms used to measure the time delays. Note that $\mathcal{T}_\mathrm{obs}$ needs to last from months to years to get a time delay measurement, however, we assume a short observing epoch, for example less than one year with high cadence~\citep{Courbin2018} that would keep the source approximately motionless in the map. For the case that light curves are monitored with many years such that precise time delays are available,
the corresponding difference should be $\int\delta\Delta t_\mathrm{micro}d\mathcal{T}_\mathrm{obs}/\int d\mathcal{T}_\mathrm{obs}$ since most algorithms try to balance each piece of
the light curves and give the average (best-fit) time delay estimation. For the long campaign case, see a companion paper~\citep{Liao2020}.

Now we can compare the measured time delay difference between bands and its theoretical distribution predicted by a disk model.
The statistic uncertainty from observation is
\begin{equation}
\sigma_{\delta \Delta t_\mathrm{lc}}=\left(\sigma^2_{\Delta t_{\mathrm{lc},b1}}+\sigma^2_{\Delta t_{\mathrm{lc},b2}}\right)^{1/2}\sim\sqrt{2}\sigma_{\Delta t_\mathrm{lc}},
\end{equation}
where $\sigma_{\Delta t_\mathrm{lc}}$ is the absolute uncertainty level of current strong lens time delay measurements based on light curves.
If it is smaller than or comparable to the theoretical distribution dispersion from quasar models, this test can become decisive with a number of measurements.
Current measurement precision has already met the requirement. We will do a simulation in the next section to show how our method works.

\section{Simulations and results}
In this section, we firstly take a PG 1115+080-like lens for example, then we discuss how to distinguish disk models with sets of time delay measurements.
A flat $\Lambda$CDM universe with $H_0=70\mathrm{km/s/Mpc}$ and the matter density parameter $\Omega_\mathrm{M}=0.3$ is assumed.
\subsection{A PG 1115+080-like lens}

\begin{figure}
\includegraphics[width=8cm,angle=0]{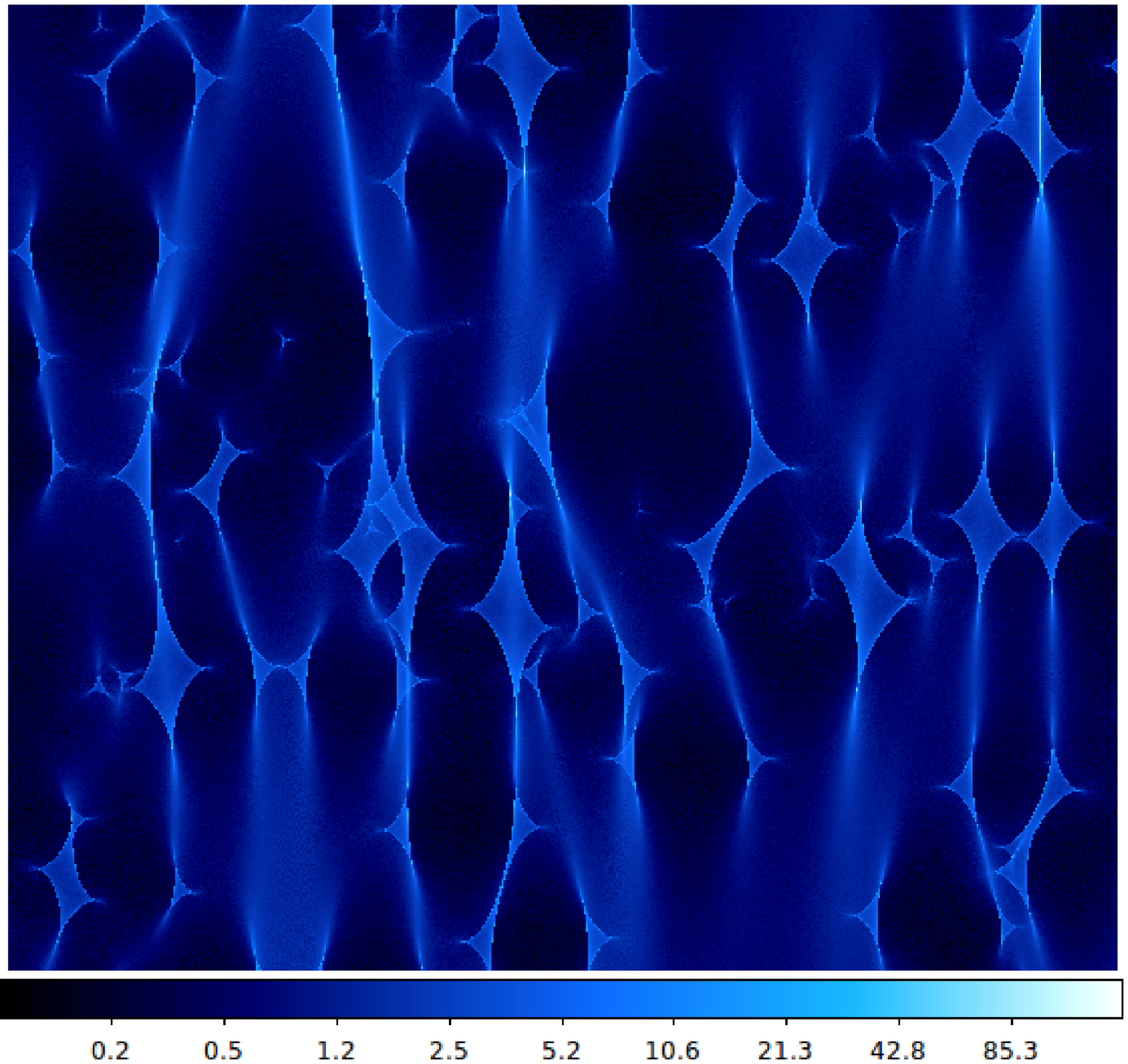} \\
\caption{Microlensing magnification map for image B. The size is $20\langle R_\mathrm{Ein}\rangle\times20\langle R_\mathrm{Ein}\rangle$ and the mean magnification equals 1.
}
\label{magB}
\end{figure}

\begin{figure}
\includegraphics[width=8cm,angle=0]{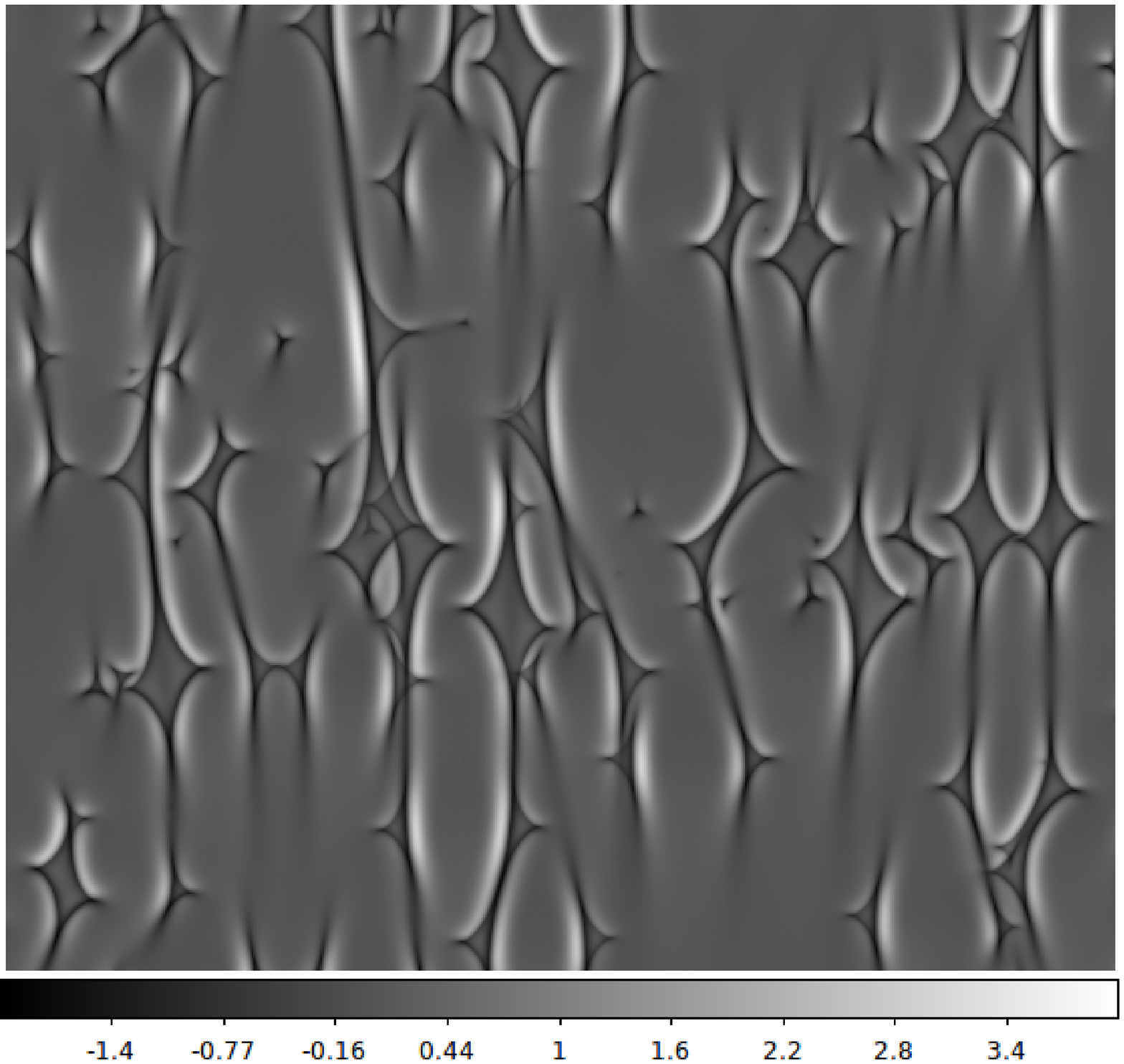} \\
\includegraphics[width=8cm,angle=0]{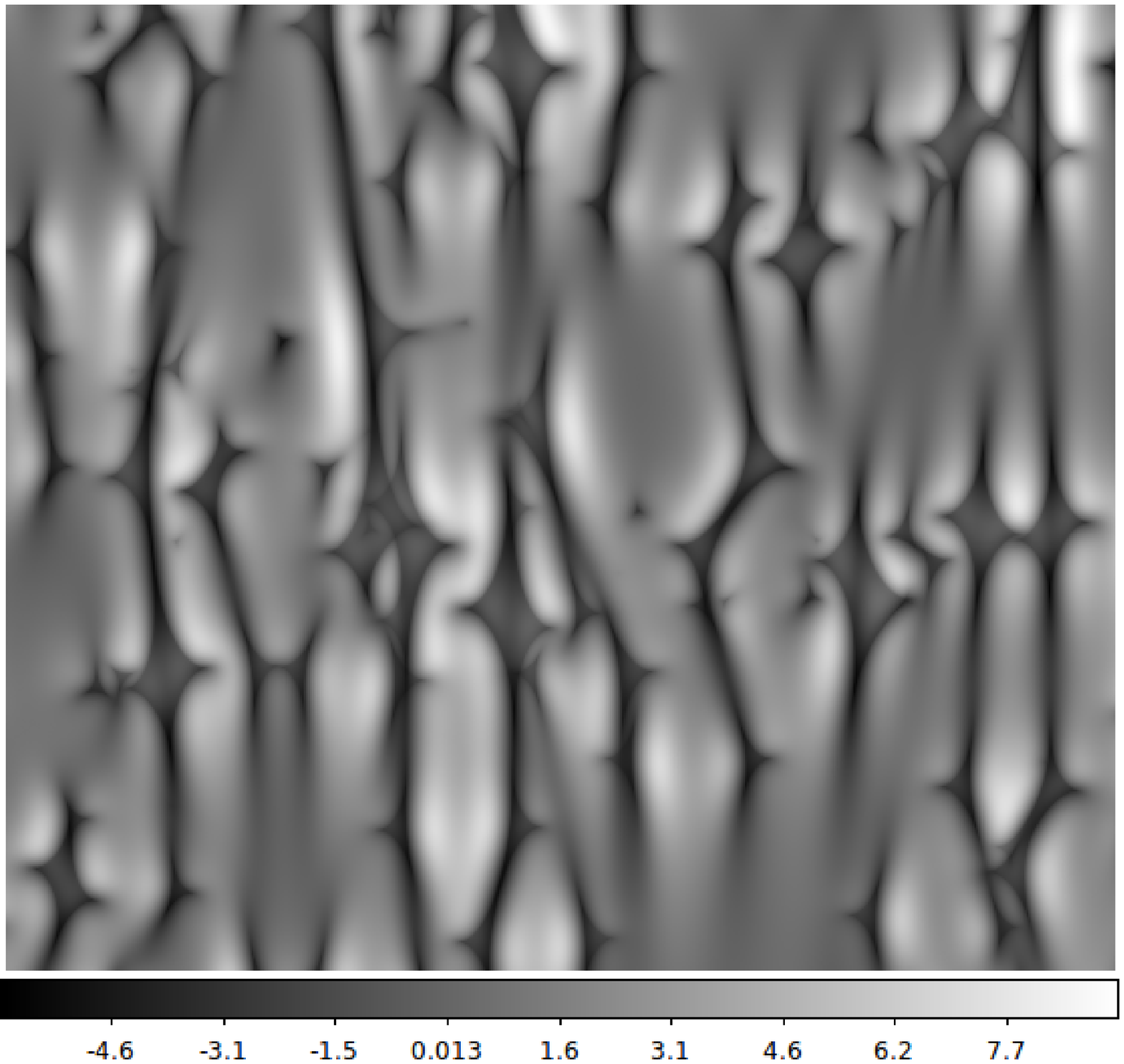}
\caption{Microlensing time lag maps in units of day of image B in the standard disk model for u band (upper panel) and i band (lower panel), respectively. $\beta=30^\circ$ and PA=$0^\circ$.
}
\label{tui}
\end{figure}

\begin{figure}
\includegraphics[width=\columnwidth,angle=0]{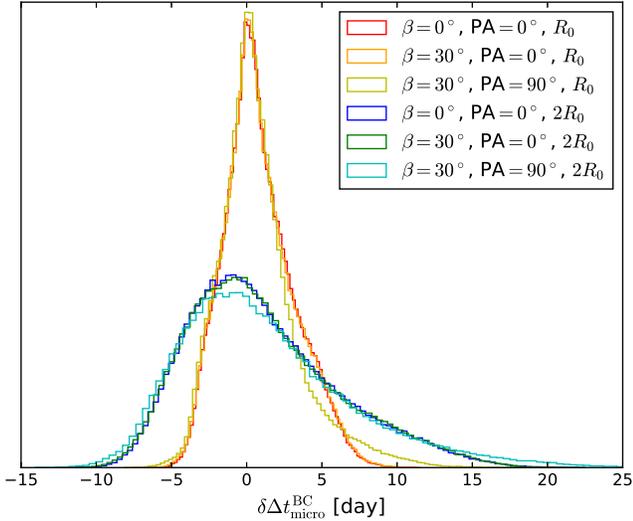} \\
\caption{Probability distributions of $\delta\Delta t^\mathrm{BC}_\mathrm{micro}$ for different cases.
}
\label{results}
\end{figure}

For illustration purpose, we take a PG 1115+080-like lens system for example, who is a quadruply imaged quasar~\citep{Weymann1980,Hege1981,Kundic1997}.
The source and lens redshifts are $z_s=1.722$ and $z_l=0.311$, respectively. The newest time delay measurements are presented in~\citep{Bonvin2018}.
Note that the two brightest images A1 and A2 are only separated by $\sim0.5$ arcseconds. Their measured fluxes are merged together in a single component simply
called image A. To avoid any issues, we adopt image B and C in this lens for the simulations. The time delay $\Delta t^\mathrm{BC}_\mathrm{lc}\sim20$ days and the parameters related with
microlensing are presented in Tab.\ref{parameters} including the local convergence $\kappa$, shear $\gamma$ and star proportion $\kappa_*/\kappa$.
These parameters come from the macro modelling of the lens and the assumption of stellar mass-luminosity ratio~\citep{Chen2019}.
\begin{table}
\centering
 \begin{tabular}{lcccc}
  \hline\hline
   Lens & Image  & $\kappa$  & $\gamma$  & $\kappa_*/\kappa$\\
  \hline
  PG 1115+080  &  B   & 0.502  &  0.811  &  0.331\\
               &  C   & 0.356  &  0.315  &  0.203\\
  \hline\hline
 \end{tabular}
 \caption{Microlensing parameters for image B and C, respectively.
}\label{parameters}
\end{table}

For the accretion disk of the source quasar, we consider a standard thin-disk model. The luminosity $L/L_\mathrm{E}$ and the efficiency $\eta$ are both set to be 0.1.
The estimated black hole mass $M_\mathrm{BH}=1.2\times10^9M_\odot$~\citep{Peng2006}. These result in $R_0(u)=7.24\times10^{14}$ cm and $R_0(i)=2.08\times10^{15}$ cm
for u band centered around 354nm and i band centered around 780nm, respectively. Furthermore, we consider a non-standard model within which the disk size is twice larger
than the standard one $R_\mathrm{disk}=2R_0$, for comparison.

We firstly generate the microlensing magnification maps~\footnote{The microlensing code used in this work, MULES, is freely available at https://github.com/gdobler/mules.} for image B and C, respectively. We assume the mean mass of the stars as the microlenses
$\langle M_*\rangle=0.3M_\odot$ and the Salpeter mass function is adopted with a ratio of the upper and lower masses being $M_\mathrm{upper}/M_\mathrm{lower}=100$.
The maps have a size of $20\langle R_\mathrm{Ein}\rangle\times20\langle R_\mathrm{Ein}\rangle$, where the mean Einstein radius $\langle R_\mathrm{Ein}\rangle=3.6\times10^{16}$cm in the source plane,
with a pixel resolution of $8192\times8192$. Note that in practice we have to generate a larger magnification map due to the integration at the edges of the map.
We use multiple maps with different random seeds to get average results.
One realization of the magnification maps of image B is shown in Fig.\ref{magB}.

Secondly, basing on Eq.\ref{tmicro}, we generate the microlensing time lag maps for the two images, the two bands and the two disk models, respectively.
The inner edge of the disk is set to be $R_\mathrm{in}=R_\mathrm{disk}/100$.
The integral radius in the disk plane is chosen to be 30$R_\mathrm{disk}$ which is large enough for including the whole disk and it is consistent with the case without microlensing.
Since the disk can not be observed, we consider three cases: 1) $\beta=0^\circ$, PA=$0^\circ$, 2) $\beta=30^\circ$, PA=$0^\circ$ and 3) $\beta=30^\circ$, PA=$90^\circ$, where
$\beta$ and PA are the inclination angle and the position angle of the disk relative to the source plane, respectively.
Note the primary factor that impacts on the results is $R_0$ and finer choices of $\beta$ and PA have little impact.
Fig.\ref{tui} shows one realization of the microlensing time lag maps of image B for band u and i, respectively, under the standard model with
$\beta=30^\circ$, PA=$0^\circ$. This figure actually implies time delay measured in the bluer band is closer to the truth than that measured in the redder band,
though the bluer band has smaller disk size that might potentially suffer from more contamination of microlensing light curves~\citep{Liao2020}.
We repeat such a process for image C whose variation in the time lag map is much smaller. Similar results can be also found in~\citep{Bonvin2018}.

Finally, we get the microlensing time delay difference between bands:
\begin{equation}\label{t1}
\delta\Delta t^\mathrm{BC}_\mathrm{micro}= \left[t^\mathrm{B}_\mathrm{micro}(i)-t^\mathrm{C}_\mathrm{micro}(i)\right]-\left[t^\mathrm{B}_\mathrm{micro}(u)-t^\mathrm{C}_\mathrm{micro}(u)\right],
\end{equation}
equivalently
\begin{equation}\label{t2}
\Delta\delta t^\mathrm{BC}_\mathrm{micro}=\left[t^\mathrm{B}_\mathrm{micro}(i)-t^\mathrm{B}_\mathrm{micro}(u)\right]-\left[t^\mathrm{C}_\mathrm{micro}(i)-t^\mathrm{C}_\mathrm{micro}(u)\right].
\end{equation}
The second expression is related with \emph{microlensing reverberation mapping difference} which measures the time lags between bands for individual images (plus $t_\mathrm{disk}(i)-t_\mathrm{disk}(u)$).
Note that this approach can be used to verify the microlensing time delay effect as well if reverberation mapping is believed to be accurate.
Comparing $\delta\Delta t_\mathrm{lc}$ and $\Delta\delta t_\mathrm{lc}$ (note the order of $\Delta$ and $\delta$ matters) can cross-check strong lens time delay and reverberation mapping measurements.

We calculate its distribution $\mathcal{P}(\delta\Delta t^\mathrm{BC}_\mathrm{micro})$ by randomly selecting total $10^6$ points in the maps of image B and C, respectively,
which have marginalized the effects of random seeds for generating magnification maps.
Given the microlensing parameters, it primarily depends on the disk size $R_\mathrm{disk}(\lambda)$. Larger $R_\mathrm{disk}$ results in
larger dispersion $\sigma_{\delta\Delta t^\mathrm{BC}_\mathrm{micro}}$.

We summarize the time delay differences for all cases in Fig.\ref{results} and Tab.\ref{statistics}.
The dispersions $\sigma_{\delta\Delta t^\mathrm{BC}_\mathrm{micro}}$ are $\sim$2.3 days and $\sim$5.0 days for the standard thin-disk model and the non-standard model with twice disk size, respectively.
Depending on the quality of the light curves, primarily the cadence, current measurement precision of time delays $\sqrt{2}\sigma_{\Delta t_\mathrm{lc}}$ could reach
$\sim$1-3 days~\citep{Courbin2018,Bonvin2019,Denzel2020}, which is comparable to/smaller than the dispersion of time delay difference induced by microlensing.
This makes such an effect detectable, especially for larger disk sizes. In fact, some evidences have shown the disk size could be even $\sim3-4$ times larger than the standard.

\begin{table*}
\centering
 \begin{tabular}{lccccccc}
  \hline\hline
   Disk size & $\beta(^\circ)$  &  PA($^\circ$)  &  Mean & 50th percentile (median)  & 16th percentile  & 84th percentile & $\sigma_{\delta\Delta t_\mathrm{micro}}$\\
  \hline
  $R_0$ & 0 & 0 &  0.837  & 0.540 &  -1.443 &  3.263  &2.353\\
  $R_0$ & 30 & 0 &  0.832 &   0.520  & -1.469  & 3.294  &2.382\\
  $R_0$ & 30 & 90 & 0.792  & 0.380  & -1.627  & 3.024  &2.326\\
  $2R_0$ & 0 & 0 & 1.070 &  0.272 &  -3.703  & 6.131  &4.917\\
  $2R_0$ & 30 & 0 & 1.069 &  0.272  & -3.755  & 6.186  &4.971\\
  $2R_0$ & 30 & 90 & 1.244 &  0.265  & -4.069  & 6.619  &5.344\\
  \hline\hline
 \end{tabular}
 \caption{Statistics of $\delta\Delta t^\mathrm{BC}_\mathrm{micro}$ in units of day for different cases, where the dispersion $\sigma_{\delta\Delta t_\mathrm{micro}}$ is defined as percentile $\mathrm{(84th-16th)}/2$.
}\label{statistics}
\end{table*}

\subsection{Statistic analysis with sets of time delays}
What one can directly measure are multiple time delay differences between bands with sets of lensing systems over the same monitoring campaign, which give the
probability distribution $\mathcal{P}(\delta\Delta t_\mathrm{lc})$.
The consistency between $\mathcal{P}(\delta\Delta t_\mathrm{micro}|R_\mathrm{disk})$ and $\mathcal{P}(\delta\Delta t_\mathrm{lc})$
would answer the question whether a quasar disk size is large enough to fit the observation.
In practice, the size of dispersion relative to the observationally statistic one $\mathcal{R}=\sigma_{\delta\Delta t_\mathrm{micro}}/(\sqrt{2}\sigma_{\Delta t_\mathrm{lc}})$ can be seen as the metric to assess
the power of our method. Larger $\mathcal{R}$ requires less lensing systems to make a conclusion.

To make our point clearer, we show it in a Bayesian framework. We define a size factor $x=R_\mathrm{disk}/R_0$, where $x=1$ corresponds to the standard model.
Suppose that $N$ time delay difference measurements are available $\delta\Delta t_\mathrm{lc}=\{\delta\Delta t^i_\mathrm{lc}\}$,
and the black hole mass and accretion efficiency parameters $M_\mathrm{BH}=\{M^i_\mathrm{BH}\}$, $\eta=\{\eta^i\}$, where $i=1,2...N$,
the posterior distribution of $x$ is given by
\begin{equation}\label{x}
\mathcal{P}(x,M_\mathrm{BH},\eta|\delta\Delta t_\mathrm{lc})\propto \mathcal{P}\left(\delta\Delta t_\mathrm{lc}|x,M_\mathrm{BH},\eta\right)
\mathcal{P}(M_\mathrm{BH})\mathcal{P}(\eta)\mathcal{P}(x),
\end{equation}
where the likelihood depends on the time lag maps and the statistic uncertainties from observations.
If the marginalized $x=1$ is excluded within some confidence levels, so is the standard thin-disk model.

Nevertheless, rather than conducting a Bayesian parameter estimation analysis,
we do a simplified simulation to show how many measurements like $\delta\Delta t^\mathrm{BC}_\mathrm{lc}$ in PG 1115+080
are required to exclude the standard model assuming the data are simulated based on $R_\mathrm{disk}=2R^\mathrm{max}_0$ for example.
$R^\mathrm{max}_0$ is the maximum value that a standard model allows by scaling the parameters $M_\mathrm{BH}$ and $\eta$ in
permitted ranges, for example, a flat prior [0.1,0.4] for $\eta$ and uncertainty $\sim0.4$ dex for black hole mass measurements by widths of different emission lines.
The black hole mass and accretion efficiency are therefore fixed in the simulation.
Considering these parameters would bring extra $\sim50\%$ uncertainty in the disk size estimation.
Note that using $R^\mathrm{max}_0$ is consistent with current works in this area which often compare the theoretically maximum disk size with the measured one.
A more detailed study should be done in the future incorporating the uncertainties of
$M_\mathrm{BH}$ and $\eta$ into consideration, which will make the constraints more stringent.
For a Bayesian analysis, one has to calculate the distributions for each $R_\mathrm{disk}$
while now we only compare two distributions as a first step for studies in this area.
We do not assume any prior on $x$ though in practice other methods like reverberation mapping and
traditional microlensing can provide inputs as a combined constraint.

For the statistic uncertainty of $\delta \Delta t_\mathrm{lc}$ by light curve shifting, we consider a constant error of 2 days following the Gaussian distribution.
We use K-S test to estimate how many measurements are required to 1) verify the microlensing time delay effect ($\delta \Delta t_\mathrm{lc}\equiv0$ without such an effect); 2) exclude the standard disk model.
Given $N$ measurements of $\delta \Delta t_\mathrm{lc}$,
we run the K-S test for 1000 times to obtain the average p-value.
We repeat this experiment with different $N$, until we find the minimal one such that the p-value$<5\%$.
We find that for the first goal, this procedure returns $N_\mathrm{min}=$4 while for the second goal
$N_\mathrm{min}=35$ which corresponds to $\sim12$ lenses for quad-image case.

\section{Summary and prospectives}
We propose an independent and robust method to verify the microlensing time delay effect which is recently known by the community and
measure the size of accretion disk using strong lens time delay differences between any two bands. The measurements are direct and
free of the issues or assumptions related with intrinsic light curves, the Hubble constant, the lens modelling, the peculia motion and the dark matter substructure.
In addition, the time delay differences are absolute quantities, therefore there is no special requirement of the lensing configurations which give different sizes of time delays.
They only depend on the absolute uncertainties of the time delay measurements based on light curves.
Our method is a supplement to the continuum reverberation mapping for studying the disk size. The latter measures the delays between bands while we measure the delays between images bypassing the assumption of
correlation between bands.

For future studies in this field, we have several comments/suggestions:

1 ) We only take bands u and i for example in this work. In fact, bands with larger wavelength difference would have more significant effects.
For example, comparing band z or y having central wavelength close to 1000nm with band u or g with shortest wavelengths will be more promising;
Surveys like LSST will get time delays in multiple bands and their differences as well, which may be correlated with each other.
Such multiple-band surveys will further strengthen the power of our method.
The required number of high-quality lenses will therefore be less. A study of the consistency among measurements in different bands will further test the details of quasar models.

2) Like in many studies of traditional microlensing, we have fixed the mean mass to be $0.3M_\odot$ while one can either use inputs from other observations or take it as an extra parameter.
Generally speaking, microlensing constrains the relative disk size to the mean Einstein radius of the stars in the lens galaxy.
The impacts of the mean stellar mass as well as the mass function shape in our method is worth further studying;

3) Distribution of microlensing time delay depends on the local environments ($\kappa,\gamma,\kappa_*/\kappa$) of the images that determine the magnification maps.
Microlensing time lag maps trace magnification maps (compare Fig.\ref{magB} and Fig.\ref{tui}) whose local-environment-dependent dispersions were discussed in the appendix of~\citep{Liao2015}.
One could choose systems with microlensing parameters and black hole masses that generate larger variations in this study whereas smaller variations are preferred in cosmological studies.

4) Compared to doubly imaged systems, quads can provide multiple time delays simultaneously which are more powerful and could be the first choice for this study;

5) Wide-field surveys like what reverberation mapping is using, for example, the Sloan Digital Sky Survey (SDSS), the PanSTARRS, the Dark Energy Survey (DES)~\citep{Jiang2017,Mudd2018} can be used in this method as well. For those whose sampling is sparse for individual bands like LSST, dedicated follow-up monitoring may be required;

Understanding the details of such a new effect would benefit understandings of quasars, microlensing and
the accuracy of strong lens time-delay cosmography that may solve the Hubble constant
issue ultimately. We look forward to seeing chromatic microlensing time delays be measured soon.

\section*{Acknowledgments}
I thank S.S. Tie, V. Bonvin and J.H.H. Chan for sharing the calculation details in their previous works and M. Sun for helpful discussions on the thin-disk model.
This work was supported by the National Natural Science Foundation of China (NSFC) No.~11973034 and
the Fundamental Research Funds for the Central Universities (WUT: 2020IB022).

\section*{Appendix}

\begin{figure}
\includegraphics[width=\columnwidth,angle=0]{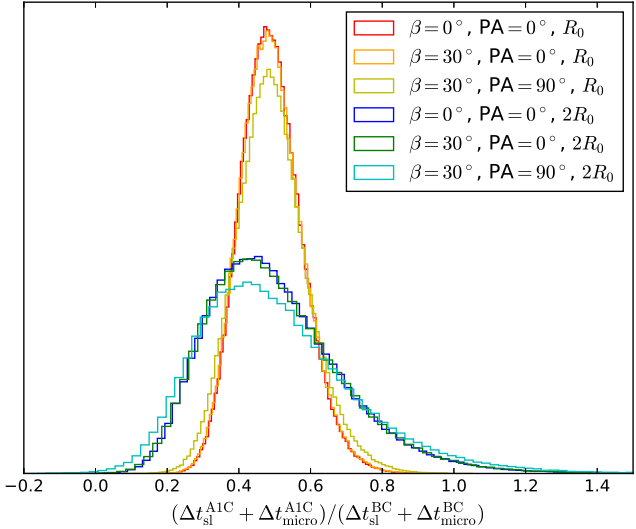}
\caption{Perturbation of time delay ratio by microlensing.
}\label{tratio}
\end{figure}

\section*{Time delay ratio anomalies}
In this appendix, we introduce using time delay ratio anomalies to test the microlensing time delay effect and quasar models.
Different from the time delay itself, this dimensionless quantity is free of various degeneracies (for example, the Hubble constant) except for dark matter substructure
perturbation that may sub-dominates. However, an unbiased lens modelling process has to be assumed such that any deviation can be attributed to microlensing rather than systematic errors.

We still take the PG 1115+080-like lensed quasar for example. We consider the time delays between image A1 and C, B and C, respectively.
For image A1, the microlensing parameters $\kappa=0.424,\ \gamma=0.491,\ \kappa_*/\kappa=0.259$.
We calculate the probability distribution of the time delay ratio $\mathcal{P}\left[(\Delta t^\mathrm{A1C}_\mathrm{sl}+\Delta t^\mathrm{A1C}_\mathrm{micro})/(\Delta t^\mathrm{BC}_\mathrm{sl}+\Delta t^\mathrm{BC}_\mathrm{micro})\right]$ measured in i band, where $\Delta t^\mathrm{A1C}_\mathrm{sl}=10$ days and
$\Delta t^\mathrm{BC}_\mathrm{sl}=20$ days, respectively, giving the benchmark ratio 0.5. The results are shown in Fig.\ref{tratio}. The relative dispersion can be up to $\sim18\%$ and $\sim38\%$ for the standard and non-standard disk sizes, respectively,
which are much larger than that inferred by lens modelling, typically at percent level.
It shows this method could be quite powerful even for a bad lens modelling.

Note that the microlensing time delays are absolute perturbations while time delay ratio inferred from lens modelling as a function of lens parameters
is a fractional estimation. The detectability $\propto 1/\Delta t_\mathrm{sl}$.  Therefore, to augment the anomalies such that they can be distinguished from statistic uncertainties, one should select
systems with shorter time delays. Remarkably, close images like A1 and A2 that have very small time delay can be identified in contrary arrival-time ordering.
It is worth pointing out that the microlensing time delay effect could be more important than dark matter substructure for detecting the anomalies.
Our work may re-open studies in this area. On the other hand, for cosmological studies, we suggest that one should check the consistency between the measured time delay ratios and those from lens modelling before inferring $H_0$.

\clearpage


\begin{thebibliography}{}
\bibitem[Birrer et al.(2016)]{Birrer2016} Birrer, S., Amara, A., \& Refregier, A. 2016, JCAP, 8, 020
\bibitem[Blackburne et al.(2011)]{Blackburne2011} Blackburne, J. A., Pooley, D., Rappaport, S., \& Schechter, P. L. 2011, ApJ, 729, 34
\bibitem[Bonvin et al.(2018)]{Bonvin2018} Bonvin, V., Chan, J. H. H., Millon, M., et al. 2018, A\&A, 616, A183
\bibitem[Bonvin et al.(2019)]{Bonvin2019} Bonvin, V., Millon, M., Chan, J. H.-H., et al. 2019, A\&A, 629, A97


\bibitem[Cackett et al.(2007)]{Cackett2007} Cackett, E. M., Horne, K., \& Winkler, H., 2007, MNRAS, 380, 669
\bibitem[Chan et al.(2020)]{Chan2020} Chan, J. H.-H., Millon, M., Bonvin, V., \& Courbin, F. 2020, A\&A, 636, A52
\bibitem[Chartas et al.(2012)]{Chartas2012} Chartas, G., Kochanek, C. S., Dai, X., et al. 2012, ApJ, 757, 137
\bibitem[Chen et al.(2018)]{Chen2018} Chen, G. C.-F., Chan, J. H. H., Bonvin, V., et al. 2018, MNRAS, 481, 1115
\bibitem[Chen et al.(2019)]{Chen2019} Chen, G. C.-F., Fassnacht, C. D., Suyu, S. H., et al. 2019, MNRAS, 490, 1743


\bibitem[Collier et al.(1998)]{Collier1998} Collier, S. J., Horne, K., Kaspi, S., et al. 1998, ApJ, 500, 162
\bibitem[Courbin et al.(2018)]{Courbin2018} Courbin, F., Bonvin, V., Buckley-Geer, E., et al. 2018, A\&A, 609, A71

\bibitem[Denzel et al.(2020)]{Denzel2020} Denzel, P., Coles, J. P., Saha, P., \& Williams, L. L. R. 2020, arXiv: 2007.14398
\bibitem[Ding et al.(2020)]{Ding2020} Ding, X., Treu, T., Birrer, S., et al. 2020, arXiv: 2006.08619

\bibitem[Edelson et al.(2015)]{Edelson2015} Edelson, R., Gelbord, J. M., Horne, K., et al. 2015, ApJ, 806, 129

\bibitem[Fabian et al.(2002)]{Fabian2002} Fabian, A. C., Vaughan, S., Nandra, K., et al. 2002, MNRAS, 335, L1
\bibitem[Fausnaugh et al.(2016)]{Fausnaugh2016} Fausnaugh, M. M., Denney, K. D., Barth, A. J., et al. 2016, ApJ, 821, 56


\bibitem[Gilman et al.(2020)]{Gilman2020} Gilman, D., Birrer, S., \& Treu, T. 2020, arXiv: 2007.01308

\bibitem[Hege et al.(1981)]{Hege1981} Hege, E. K., Hubbard, E. N., Strittmatter, P. A., \& Worden, S. P. 1981, ApJ, 248, L1

\bibitem[Iwasawa et al.(1996)]{Iwasawa1996} Iwasawa, K., Fabian, A. C., Reynolds, C. S., et al. 1996, MNRAS, 282, 1038
\bibitem[Iwasawa et al.(2004)]{Iwasawa2004} Iwasawa, K., Lee, J. C., Young, A. J., Reynolds, C. S., \& Fabian, A. C. 2004, MNRAS, 347, 411

\bibitem[Jiang et al.(2017)]{Jiang2017} Jiang, Y.-F., Green, P. J., Greene, J. E., et al. 2017, ApJ, 836, 186



\bibitem[Keeton \& Moustakas(2009)]{Keeton2009} Keeton, C. R., \& Moustakas, L. A., 2009, ApJ, 699, 1720
\bibitem[Kundic et al.(1997)]{Kundic1997} Kundic, T., Cohen, J. G.,  Blandford, R. D., \& Lubin, L. M. 1997, AJ, 114, 507
\bibitem[Liao et. al.(2015)]{Liao2015} Liao, K., Treu, T., Marshall, P., et. al. 2015, ApJ, 800, 11
\bibitem[Liao et al.(2018)]{Liao2018} Liao, K., Ding, X., Biesiada, M., Fan, X.-L., \& Zhu, Z.-H. 2018, ApJ, 867, 69
\bibitem[Liao(2019)]{Liao2019} Liao, K. 2019, ApJ, 871, 113
\bibitem[Liao(2020)]{Liao2020} Liao, K. 2020, arXiv: 2007.13996

\bibitem[Lira et al.(2015)]{Lira2015} Lira, P., Ar\'{e}valo, P., Uttley, P., McHardy, I. M. M., \& Videla, L. 2015, MNRAS, 454, 368

\bibitem[MacLeod et al.(2015)]{MacLeod2015} MacLeod, C. L., Morgan, C. W., Mosquera, A., et al. 2015, ApJ, 806, 258
\bibitem[Morgan et al.(2008)]{Morgan2008} Morgan, C. W., Kochanek, C. S., Dai, X., Morgan, N. D., \& Falco, E. E. 2008, ApJ, 689, 755
\bibitem[Morgan et al.(2010)]{Morgan2010} Morgan, C. W., Kochanek, C. S., Morgan, N. D., \& Falco, E. E. 2010, ApJ, 712, 1129
\bibitem[Mosquera et al.(2013)]{Mosquera2013} Mosquera, A. M., Kochanek, C. S., Chen, B., et al. 2013, ApJ, 769, 53
\bibitem[Mudd et al.(2018)]{Mudd2018} Mudd, D., Martini, P., Zu, Y., et al. 2018, ApJ, 862, 123


\bibitem[Oguri \& Marshall(2010)]{Oguri2010} Oguri, M., \& Marshall, P. J., 2010, MNRAS, 405, 2579
\bibitem[Peng et al.(2006)]{Peng2006} Peng, C. Y., Impey, C. D., Rix, H.-W., et al. 2006, ApJ, 649, 616
\bibitem[Peterson et al.(1998)]{Peterson1998} Peterson, B. M., Wanders, I., Horne, K., et al. 1998, Publications of the Astronomical Society of the Pacific, 110, 660
\bibitem[Pooley et al.(2007)]{Pooley2007} Pooley, D., Blackburne, J. A., Rappaport, S., \& Schechter, P. L. 2007, ApJ, 661, 19


\bibitem[Rojas et al.(2014)]{Rojas2014} Rojas, K., Motta, V., Mediavilla, E., et al. 2014, ApJ, 797, 61

\bibitem[Schneider \& Sluse(2013)]{Schneider2013} Schneider, P., \& Sluse, D. 2013, A\&A, 559, A37
\bibitem[Sergeev et al.(2005)]{Sergeev2005} Sergeev, S. G., Doroshenko, V. T., Golubinskiy, Y. V., Merkulova, N. I., \& Sergeeva, E. A. 2005, ApJ, 622, 129
\bibitem[Shakura \& Sunyaev(1973)]{Shakura1973} Shakura, N. I., \& Sunyaev, R. A., 1973, A\&A, 24, 337
\bibitem[Shappee et al.(2014)]{Shappee2014} Shappee, B. J., Prieto, J. L., Grupe, D., et al. 2014, ApJ, 788, 48
\bibitem[Sun et al.(2014)]{Sun2014} Sun, Y.-H., Wang, J.-X., Chen, X.-Y., \& Zheng, Z.-Y. 2014, ApJ, 792, 54
\bibitem[Tanaka et al.(1995)]{Tanaka1995} Tanaka, Y., Nandra, K., Fabian, A. C., et al. 1995, Nature, 375, 659
\bibitem[Tie \& Kochanek(2018)]{TK18} Tie, S. S., \& Kochanek, C. S., 2018, MNRAS, 473, 80
\bibitem[Treu(2010)]{Treu2010} Treu, T. 2010, Annu. Rev. Astron. Astrophys., 48, 87
\bibitem[Treu \& Marshall(2016)]{Treu2016} Treu, T., \& Marshall, P. J. 2016, A\&ARv, 24, 11
\bibitem[Weymann et al.(1980)]{Weymann1980} Weymann, R. J., Latham, D., Roger, J., et al. 1980, Nature, 285, 641
\bibitem[Wong et al.(2020)]{Wong2020} Wong, K. C., Suyu, S. H., Chen, G. C.-F., et al. 2020, arXiv:1907.04869
\bibitem[Zu et al.(2013)]{Zu2013} Zu, Y., Kochaneck, C. S., Koz{\l}owski, S., \& Udalski, A. 2013, ApJ, 765, 106



















\end{thebibliography}
\end{document}